\documentclass{article}

\def\xxinput#1{\input#1}

\xxinput{vsolj02.sty}

\usepackage[dvipdfmx]{graphicx}
\usepackage[comma,colon]{natbib}

\usepackage[OT2,T1]{fontenc}

\def\cite{\citealt}
\setcitestyle{aysep={}}

\newcounter{author}
\def\altaffilmark#1{$^{#1}$}
\def\altaffiltext#1{$^{#1}$\,}

\def\authorcount#1#2{{\refstepcounter{author}\label{#1}
                     \altaffiltext{\ref{#1}}{#2}}}

\begin{document}

\begin{center}

\title{MASTER OT J055845.55$+$391533.4: SU UMa star with a dip and long rebrightening}

\author{
        Taichi~Kato\altaffilmark{\ref{affil:Kyoto}},
        Hiroshi~Itoh\altaffilmark{\ref{affil:Ioh}},
        Tonny~Vanmunster\altaffilmark{\ref{affil:Vanmunster}},
        Seiichiro~Kiyota\altaffilmark{\ref{affil:Kis}},
        Katsuaki~Kubodera\altaffilmark{\ref{affil:Kub}},
}
\author{
        Pavol~A.~Dubovsky\altaffilmark{\ref{affil:Dubovsky}},
        Igor~Kudzej\altaffilmark{\ref{affil:Dubovsky}},
        Tom\'a\v{s}~Medulka\altaffilmark{\ref{affil:Dubovsky}},
        Filipp~D.~Romanov\altaffilmark{\ref{affil:Romanov}}$^,$\altaffilmark{\ref{affil:AAVSO}},
        David~J.~Lane\altaffilmark{\ref{affil:AbbeyRidge}}$^,$\altaffilmark{\ref{affil:AAVSO}}
}

\authorcount{affil:Kyoto}{
     Department of Astronomy, Kyoto University, Sakyo-ku,
     Kyoto 606-8502, Japan \\
     \textit{tkato@kusastro.kyoto-u.ac.jp}
}

\authorcount{affil:Ioh}{
     Variable Star Observers League in Japan (VSOLJ),
     1001-105 Nishiterakata, Hachioji, Tokyo 192-0153, Japan \\
     \textit{pxb02072@nifty.com}
}

\authorcount{affil:Vanmunster}{
     Center for Backyard Astrophysics Belgium, Walhostraat 1A,
     B-3401 Landen, Belgium \\
     \textit{tonny.vanmunster@gmail.com}
}

\authorcount{affil:Kis}{
     VSOLJ, 7-1 Kitahatsutomi, Kamagaya, Chiba 273-0126, Japan \\
     \textit{skiyotax@gmail.com}
}

\authorcount{affil:Kub}{
     VSOLJ, 2708-1 Kozu, Odawara, Kanagawa, 256-0812, Japan \\
     \textit{hfd00771@nifty.com}
}

\authorcount{affil:Dubovsky}{
     Vihorlat Observatory, Mierova 4, 06601 Humenne, Slovakia \\
     \textit{var@kozmos.sk}
}

\authorcount{affil:Romanov}{
     Pobedy street, house 7, flat 60, Yuzhno-Morskoy, Nakhodka, Primorsky
     Krai 692954, Russia, \\
     and
     remote observer of Abbey Ridge Observatory$^{\textrm{\ref{affil:AbbeyRidge}}}$, \\
     \textit{filipp.romanov.27.04.1997@gmail.com},
     https://orcid.org/0000-0002-5268-7735
}

\authorcount{affil:AAVSO}{
     American Association of Variable Star Observers (AAVSO)
}

\authorcount{affil:AbbeyRidge}{
     Abbey Ridge Observatory, 45 Abbey Rd, Stillwater Lake, NS,
     B3Z1R1, Canada, \\
     \textit{dave@davelane.ca},
     https://orcid.org/0000-0002-6097-8719
}

\end{center}

\begin{abstract}
\xxinput{abst.inc}
\end{abstract}

\section{Introduction}\label{sec:intro}

   MASTER OT J055845.55$+$391533.4 was detected as an optical
transient on 2014 February 19 at a magnitude of 14.4
\citep{yec14j0558atel5905}.  The object was found to be already
at 13.9 mag on 2014 February 13.  Two past outbursts had been
detected \citep{yec14j0558atel5905}.  This object was confirmed
to be an SU UMa star by the detection of superhumps in 2014
\citep{Pdot7}.  The superhump period and supercycle were suggested
to be 0.0563(4)~d and 360--450~d, respectively \citep{Pdot7}
[for general information of cataclysmic variables (CVs)
and dwarf novae (DNe), see e.g., \citet{war95book}].
\citet{Pdot9} observed this object on three nights in 2016
and obtained a period of 0.0581~d.  \citet{Pdot9} already
reported that the outburst behavior was rather strange.
We have noticed that this object showed a dip and rebrightening
in superoutbursts recorded in modern survey data (see below).
Furthermore, such superoutburst with a dip and (sometimes complex)
rebrightening occur successively without intervening normal
outbursts.  Such behavior was very unusual for a hydrogen-rich
DN but is frequently seen in helium-rich AM CVn objects
[although not very apparent in figure 2 of \citet{lev15amcvn},
CP Eri is such an object (see section \ref{sec:dis})].
We therefore initiated a time-resolved
photometic campaign during the 2023 February--March superoutburst
to see the development of superhumps and the superoutburst.
This object is also known as a variable star ZTF J055845.48$+$391533.1
\citep{ofe20ZTFvar} and Gaia DR3 3458275544681382912
(=Gaia16bgq, type CV) \citep{GaiaDR3}.

\section{Data analysis}

   We used Asteroid Terrestrial-impact
Last Alert System (ATLAS: \cite{ATLAS}) forced photometry
\citep{shi21ALTASforced}, Zwicky Transient Facility
(ZTF: \cite{ZTF})\footnote{
   The ZTF data can be obtained from IRSA
$<$https://irsa.ipac.caltech.edu/Missions/ztf.html$>$
using the interface
$<$https://irsa.ipac.caltech.edu/docs/program\_interface/ztf\_api.html$>$
or using a wrapper of the above IRSA API
$<$https://github.com/MickaelRigault/ztfquery$>$.
} and All-Sky Automated Survey for Supernovae
(ASAS-SN: \cite{ASASSN}) Sky Patrol \citep{koc17ASASSNLC}
data to examine the long-term behavior.
The ASAS-SN positive detections fainter than $V$=16.3 or
$g$=16.5 were excluded as noises close to the detection limit.
The ASAS-SN data around BJD 2459252 were included even below
this limit since the reality of these data were confirmed
by a comparison with the ZTF data.
Some unfiltered snapshot observations reported to VSOLJ
and VSNET \citep{VSNET} were also included (hereafter CCD).

   Time-resolved photometry during the 2023 February--March
superoutburst was obtained by VSOLJ members and
the VSNET Collaboration using 30cm-class telescopes.
We also obtained observations during the 2021 January--February
superoutburst.  The log of these observations is listed
in table \ref{tab:log}.
Superhumps maxima were determined using
the template fitting method introduced in \citet{Pdot}
after correcting zero-point differences between the observers
and removing outburst trends by locally-weighted polynomial
regression (LOWESS: \cite{LOWESS}).
The superhump period was determined using the phase dispersion
minimization (PDM: \cite{PDM}) method, whose errors were
estimated by the methods of \citet{fer89error,Pdot2}.

\xxinput{obs.inc}

\section{Results}

\subsection{Long-term behavior}

\begin{figure*}
\begin{center}
\includegraphics[width=15cm]{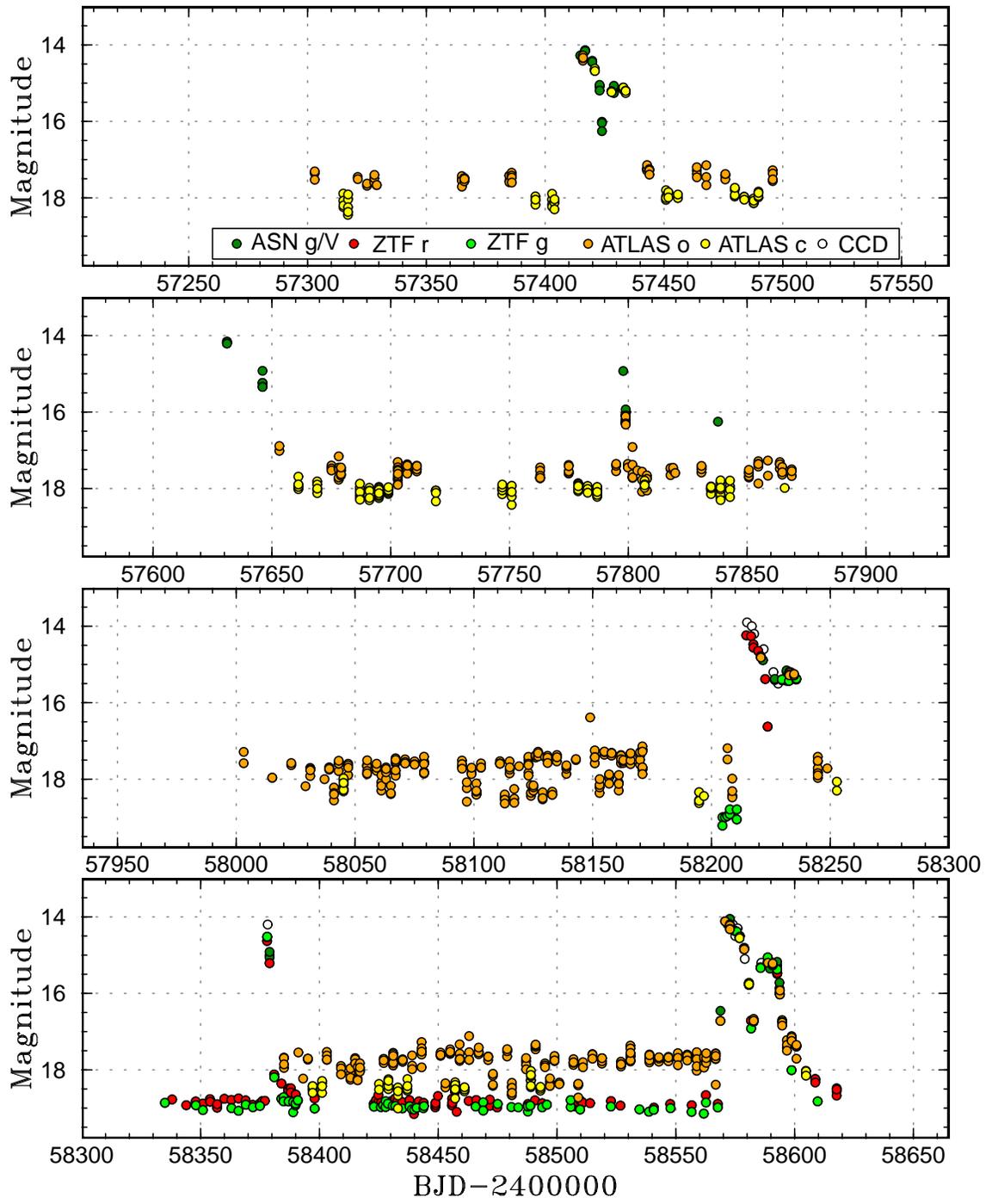}
\caption{
   Light curve of MASTER OT J055845.55$+$391533.4 in 2015--2019.
ASN refers to ASAS-SN and CCD refers to unfiltered snapshot observations.
}
\label{fig:lc1}
\end{center}
\end{figure*}

\begin{figure*}
\begin{center}
\includegraphics[width=15cm]{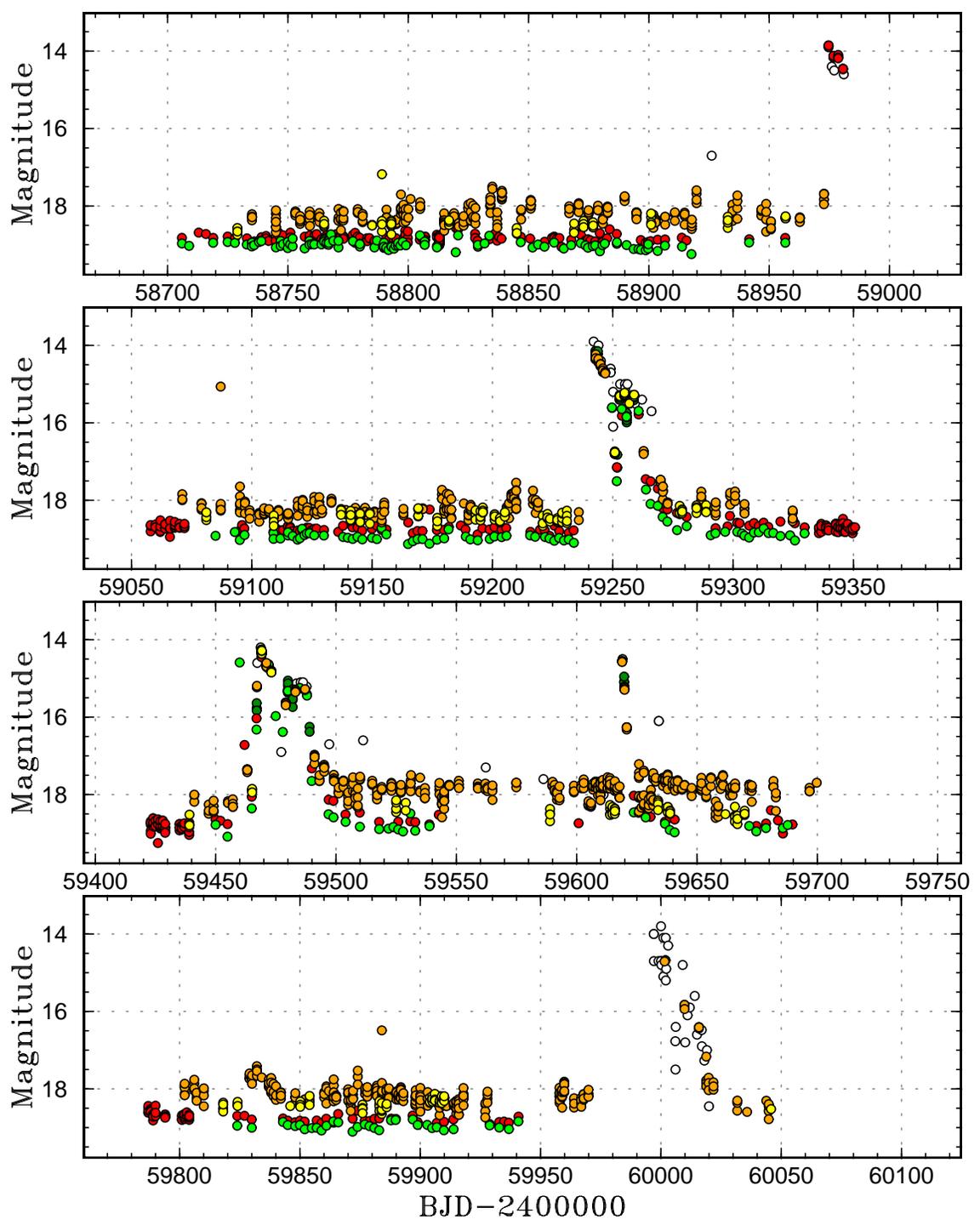}
\caption{
   Light curve of MASTER OT J055845.55$+$391533.4 in 2019--2023.
The symbols are the same as in figure \ref{fig:lc1}.
}
\label{fig:lc2}
\end{center}
\end{figure*}

\begin{figure*}
\begin{center}
\includegraphics[width=15cm]{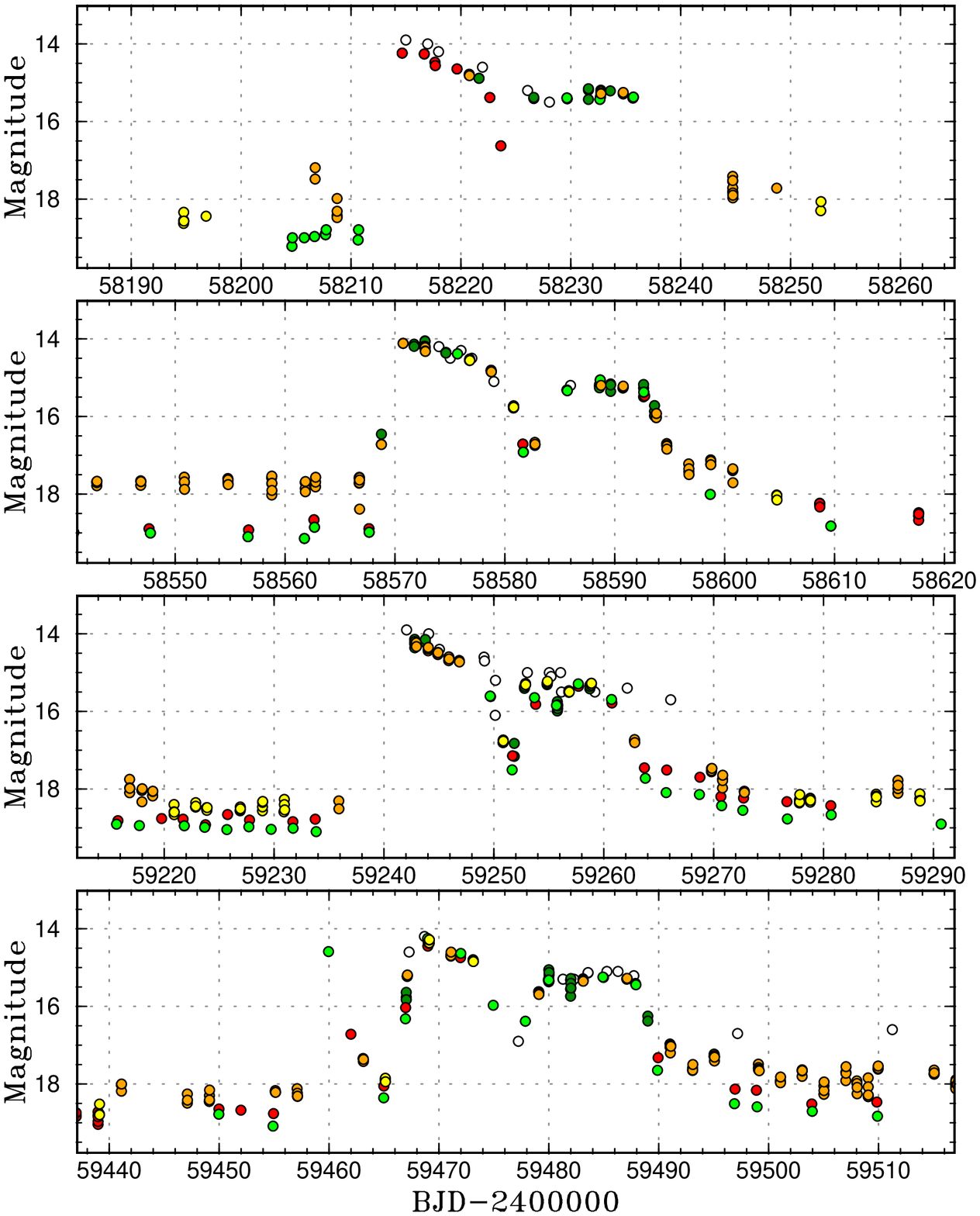}
\caption{
   Well-observed superoutbursts of 
MASTER OT J055845.55$+$391533.4.
The symbols are the same as in figure \ref{fig:lc1}.
All superoutbursts had a dip in the middle and subsequent
long rebrightening.
}
\label{fig:lc3}
\end{center}
\end{figure*}

   Long-term light curves are shown in figures \ref{fig:lc1}
and \ref{fig:lc2}.  The ATLAS data in quiescence were
systematically brighter than the ZTF data.  This is probably
due to the contamination from a red nearby (physically unrelated)
companion Gaia DR3 3458275544681383552 (Gaia $BP$=18.53 and
$RP$=16.65, \cite{GaiaDR3}).
Eight superoutbursts were recorded in this interval.
Four representative well-observed superoutbursts are also
shown in figure \ref{fig:lc3} to show more details.
\begin{itemize}
\item
2016 January--February (BJD 2457416--2457434,
first panel in figure \ref{fig:lc1}).
ASAS-SN recorded a dip in the middle of this outburst.
\item
2016 August--September (BJD 2457631--2457646,
first panel in figure \ref{fig:lc1}).
This outburst was observed by ASAS-SN with on three nights.
There was fading in the middle of the outburst,
although observations before and after it were very sparse.
There were also time-resolved observations in the late stage
\citep{Pdot9} and the long duration of the outburst was
secure. We consider that this outburst was likely a superoutburst
with a dip.
\item
2018 April--May (BJD 2458215--2458236,
third panel in figure \ref{fig:lc1} and
first panel in figure \ref{fig:lc3}).
ZTF and ASAS-SN observations clearly showed the presence of a dip
in the middle of this outburst (the ASAS-SN observation during
the dip completely overlapped the ZTF point and is invisible
in the figures).  The duration of the dip appeared to be
relatively short.
\item
2019 March--April (BJD 2458571--2458596,
fourth panel in figure \ref{fig:lc1} and
second panel in figure \ref{fig:lc3}).
A relatively long dip was clearly present in ZTF, ATLAS
and ASAS-SN observations.
\item
2020 May (BJD 2458975--2458981,
first panel in figure \ref{fig:lc2}).
Only the initial part of the outburst was recorded due to
the seasonal gap in observation.  The duration of the outburst,
however, was sufficient for a superoutburst.
\item
2021 January--February (BJD 2459242--2459263,
second panel in figure \ref{fig:lc2} and
third panel in figure \ref{fig:lc3}).
A relatively long dip was clearly present in ZTF, ATLAS
and ASAS-SN observations.  This outburst showed some short-term
variations (oscillations) after the dip.
\item
2021 September--October (BJD 2459467--2459488,
third panel in figure \ref{fig:lc2} and
fourth panel in figure \ref{fig:lc3}).
A relatively long dip was clearly present in ZTF, ATLAS and
CCD observations.  An outburst in ZTF and ATLAS on
2021 September 2--5 (BJD 2459460--2459463) was apparently
a precursor.
\item
2023 February--March (BJD 2459997--2460020,
fourth panel in figure \ref{fig:lc2}).
There was a dip in the middle of the outburst.
This is the superoutburst during which we made a time-resolved
photometric campaign in this paper (see figure \ref{fig:humpall}).
\end{itemize}

   The shortest intervals between the superoutburst was
215~d, 225~d being the second shortest.  Assuming that
two superoutbursts were missed during the seasonal gaps in
2017 and 2022, the mean interval of the superoutbursts
was 298(8)~d.

   Only three normal outbursts were detected: 2017 February 13
(BJD 2457798, second panel in figure \ref{fig:lc1}),
2018 September 16 (BJD 2458378, fourth panel in
figure \ref{fig:lc1}) and 2022 February 8 (BJD 2459619,
third panel in figure \ref{fig:lc1}).  These results indicate
that normal outbursts are essentially rare in this object.

\subsection{2023 February--March superoutburst}

\begin{figure*}
\begin{center}
\includegraphics[width=14cm]{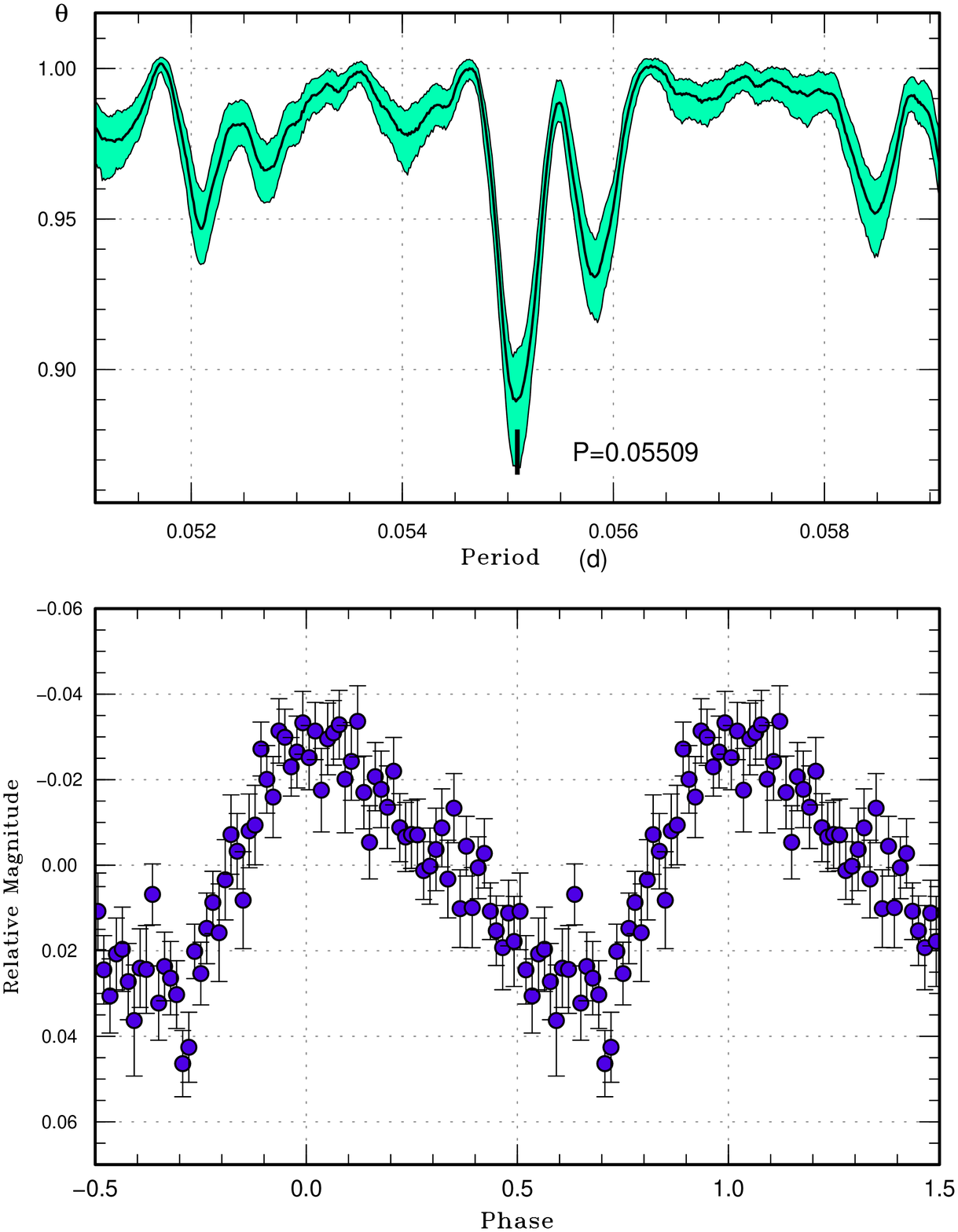}
\caption{
   Mean superhump profile of MASTER OT J055845.55$+$391533.4
   during the 2023 February--March superoutburst.
   The data before the dip (before BJD 2460005) were used.
   (Upper): PDM analysis.  The bootstrap result using
   randomly contain 50\% of observations is shown as
   a form of 90\% confidence intervals in the resultant 
   $\theta$ statistics.
   (Lower): Phase plot.
}
\label{fig:pdm}
\end{center}
\end{figure*}

\begin{figure*}
\begin{center}
\includegraphics[width=15cm]{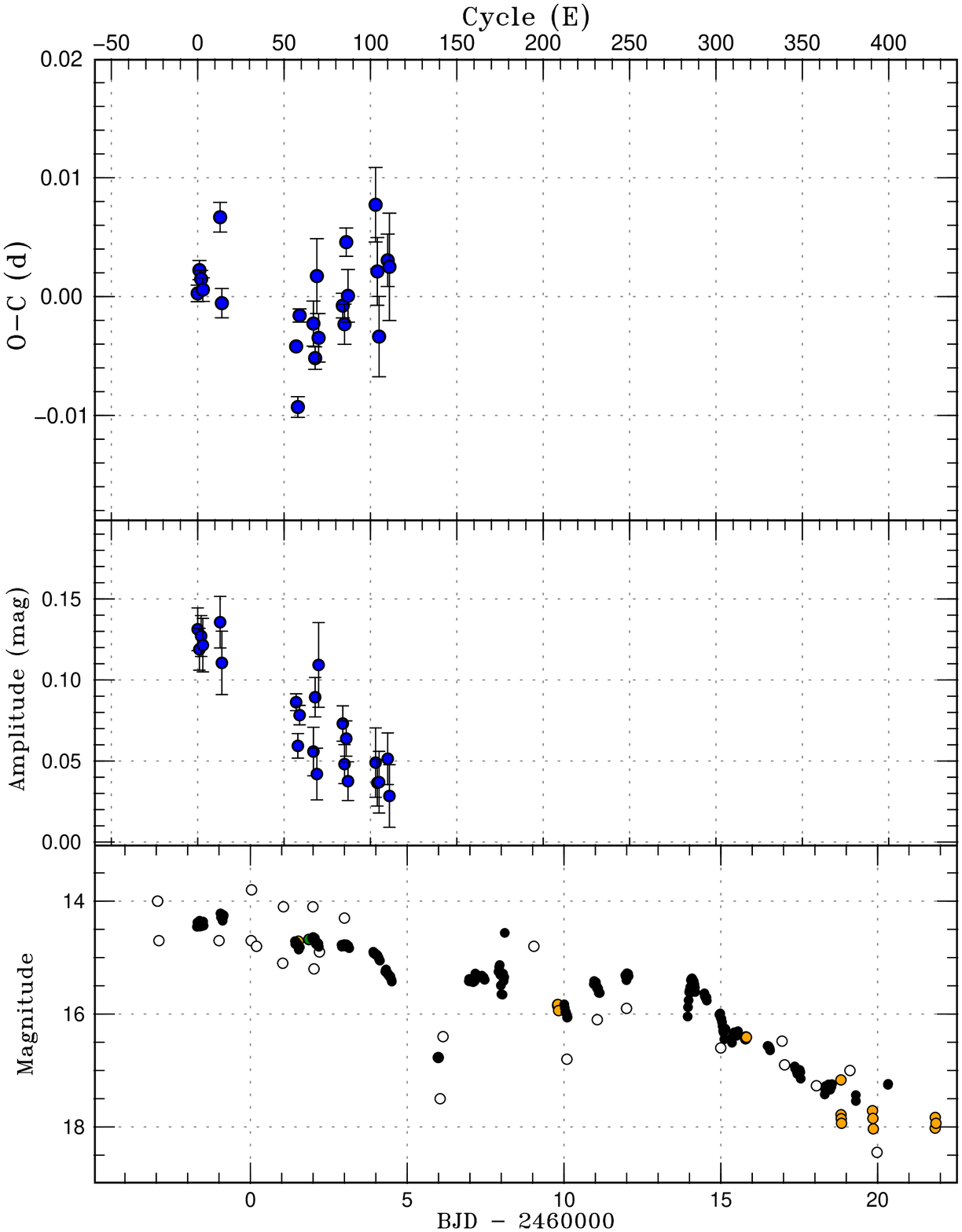}
\caption{
  2023 February--March superoutburst of
  MASTER OT J055845.55$+$391533.4.
  (Upper:) $O-C$ variation.  The ephemeris of
  BJD(max) = 2459998.3205$+$0.055079$E$ was used.
  The data are in table \ref{tab:oc}.
  (Middle:) Superhump amplitude.
  (Lower:) Light curve.
  The data were binned to 0.018~d (black filled circles).
  Other symbols are the same as in figure \ref{fig:lc1}.
}
\label{fig:humpall}
\end{center}
\end{figure*}

   A PDM analysis and the mean profile of superhumps before
the dip are shown in figure \ref{fig:pdm}.
The mean period in this interval was 0.05509(2)~d.
Superhumps were below the detection limit during
the rebrightening phase after the dip.
The times of superhump maxima are listed in table
\ref{tab:oc}.  The period derivative
($P_{\rm dot} = \dot{P}/P$) was $+$7(2) $\times$ 10$^{-5}$
(see \cite{Pdot}).
The $O-C$ diagram, variation of the superhump amplitudes and
the light curve are shown in figure \ref{fig:humpall}.

\xxinput{oc.inc}

\subsection{2021 January--February superoutburst}\label{sec:obs2021}

\begin{figure*}
\begin{center}
\includegraphics[width=14cm]{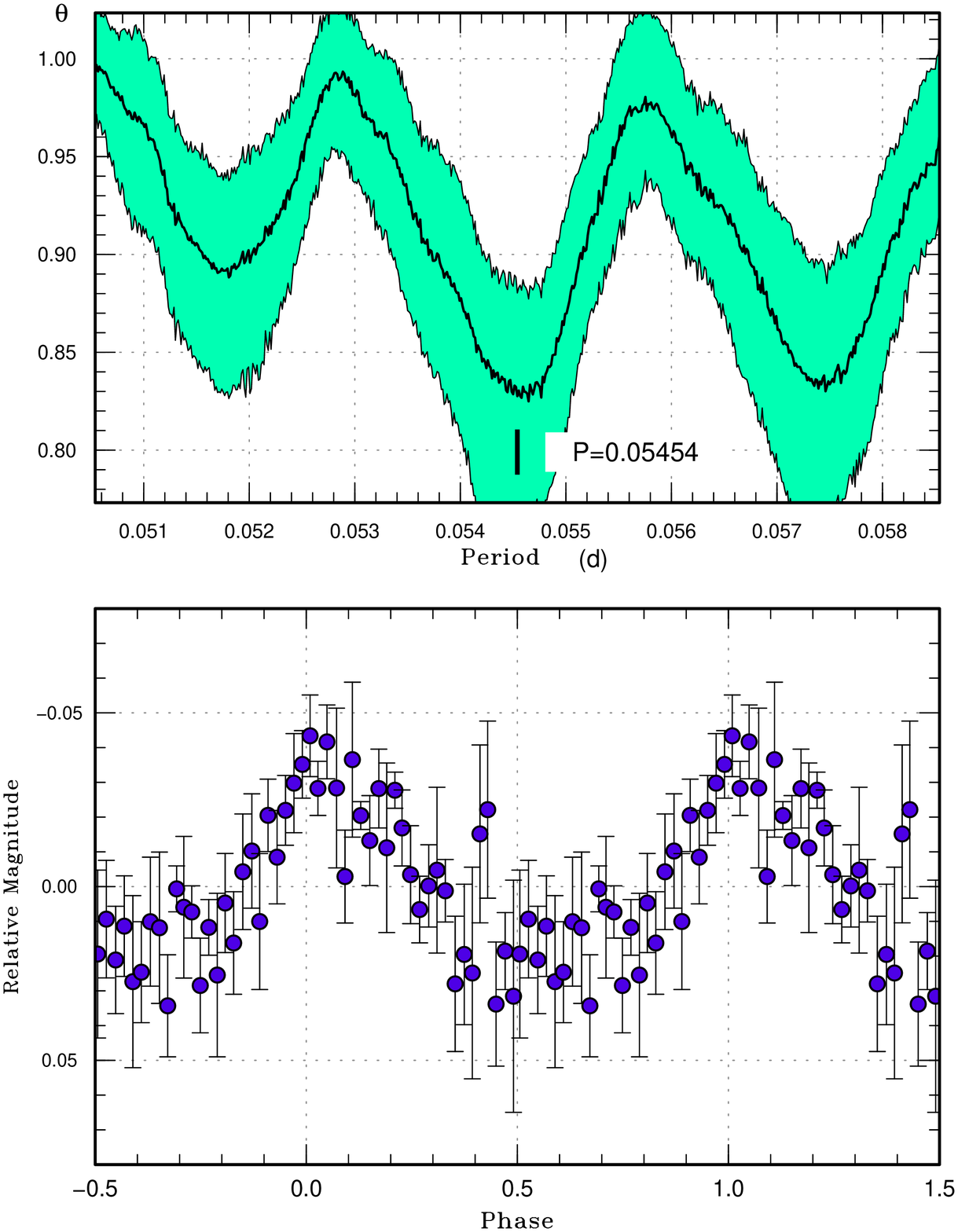}
\caption{
   Mean superhump profile of MASTER OT J055845.55$+$391533.4
   during the 2021 January--February superoutburst.
   The two-night data were obtained after the dip.
   (Upper): PDM analysis.
   (Lower): Phase plot.
}
\label{fig:pdm2021}
\end{center}
\end{figure*}

   We also observed this object on two nights during the 2021
January--February superoutburst
(second panel in figure \ref{fig:lc2}).
The observations (BJD 2459254.907--2459256.183) were after
the dip.  A PDM analysis detected superhumps with a period
of 0.05454(12)~d (figure \ref{fig:pdm2021}).
   
\section{Discussion}\label{sec:dis}

   The recorded superhumps in 2023 were identified as stage B
(for superhump stages, see \cite{Pdot}) based on the profile,
decreasing amplitude and the epoch when superhumps were
observed.  The observed $P_{\rm dot}$ = $+$7(2) $\times$ 10$^{-5}$
was not unusual for this short superhump period \citep{Pdot}.

   In \citet{Pdot7}, the 2014 superoutburst was observed
already at least 15~d after the start of the outburst and
the recorded superhumps were suggested to be stage C.
The 2023 data, however, showed no sign of
stage C nor evidence for superhumps in the late stage
of the superoutburst.  Although the period 0.0563(4)~d was
slightly different from the current measurement, it can be
consistent with the present result considering that
the amplitude of superhumps was very small in the 2014 data
and that the period was based on single-night data
by a single observer.
The 2016 observations \citep{Pdot9} were also obtained near
the end of the superoutburst.  The period (0.0581~d)
appears to be a one-day alias of the present result.
Although the 2021 observations (subsection \ref{sec:obs2021})
were limited, superhumps after the dip appeared to
have been detected. 
These results suggest that low-amplitude superhumps
were present during the rebrightening phase, which were
below the detection limit of the 2023 observations.
The superhump period during the rebrightening phase, however,
has not yet been well-determined.

\begin{figure*}
\begin{center}
\includegraphics[width=15cm]{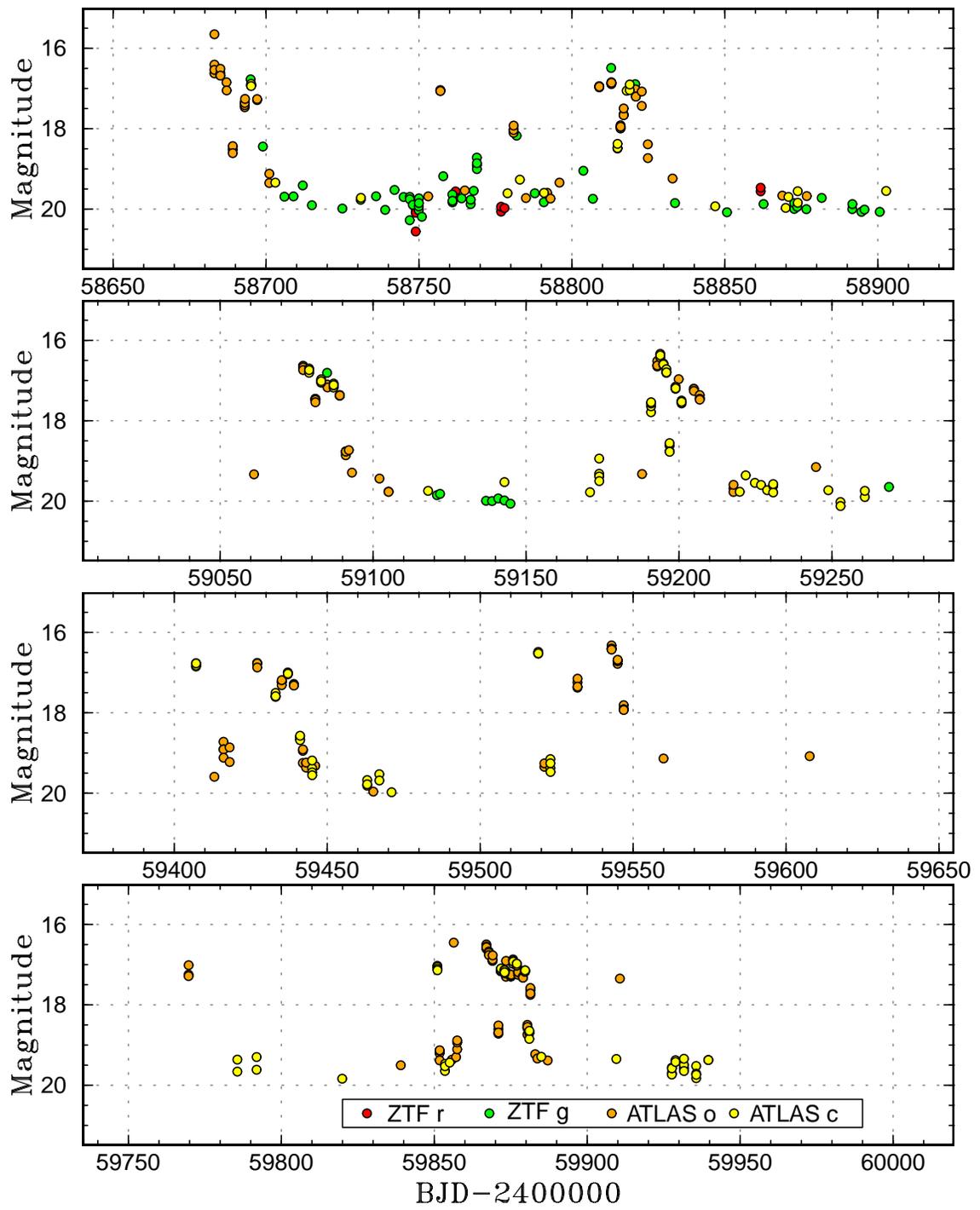}
\caption{
   Light curve of CP Eri.  The presence of a dip in the middle
of a superoutburst is the feature common to
MASTER OT J055845.55$+$391533.4.
}
\label{fig:cperi-lc1}
\end{center}
\end{figure*}

   The presence of a dip in the middle of a superoutburst
is a strange feature in this object.  As stated in
section~\ref{sec:intro}, AM CVn stars generally show this
feature.  We present a light curve of CP Eri in 
figure \ref{fig:cperi-lc1} for a comparison.
All well-observed superoutbursts have a dip in the middle
and the number of normal outbursts is relatively low
despite the short supercycle.  The observed superhump period
of MASTER OT J055845.55$+$391533.4, however, is strongly
against the possibility of an AM CVn-type object.

\begin{figure*}
\begin{center}
\includegraphics[width=15cm]{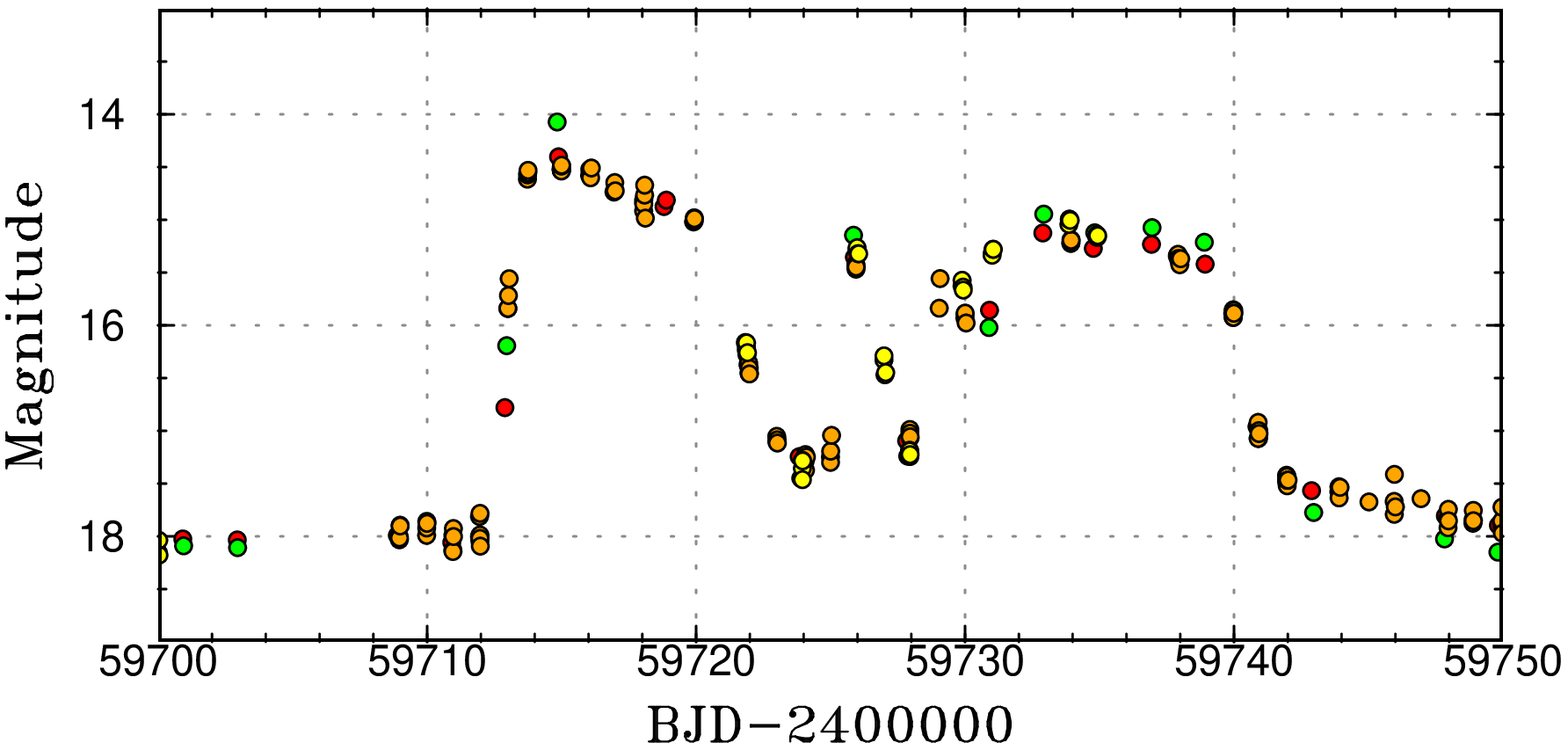}
\caption{
   Complex superoutburst in MASTER OT J172758.09$+$380021.5
recorded in 2022.
The symbols are the same as in figure \ref{fig:cperi-lc1}.
}
\label{fig:j1727-lc1}
\end{center}
\end{figure*}

   There is another known example of very complex superoutbursts
in the very unusual object MASTER OT J172758.09$+$380021.5,
which has a short superhump period of 0.05829~d and a short
supercycle of 50--100~d \citep{pav21j1727}.
The complex superoutburst recorded in 2022 is shown in
figure \ref{fig:j1727-lc1} (see also figure 5 in
\cite{pav21j1727} for the past ones).  After a relatively long dip and
short rebrightening, this object entered long rebrightening.
The behavior is similar to MASTER OT J055845.55$+$391533.4
in that the superoutburst showed a dip (but more structured)
in the middle of a superoutburst.  The short orbital period
and supercycle are also similar to MASTER OT J055845.55$+$391533.4.
In the case of MASTER OT J172758.09$+$380021.5, optical spectra
confirmed its hydrogen-rich nature \citep{tho16CVs,pav21j1727}.
The differences between MASTER OT J055845.55$+$391533.4 and
MASTER OT J172758.09$+$380021.5 are that the dip in superoutburst
is highly reproducible in the former while it is not in the latter
and that the supercycle is a few times shorter in the latter.

   Such a dip during a superoutburst and long rebrightening
are usually seen in WZ Sge stars \citep{kat15wzsge},
but not in ordinary SU UMa stars.
In WZ Sge stars, a cooling wave in the accretion disk
somehow occurs during a superoutburst and the remaining matter
in the disk is considered to be responsible for long or
repeated rebrightening \citep{mey15suumareb}.
Low mass-ratios ($q$) such as in WZ Sge stars
(and some ER~UMa stars such as RZ LMi) have been proposed to be
a cause of such premature quenching of a superoutburst
\citep{hel01eruma,osa95rzlmi}.
The $q$ value of MASTER OT J172758.09$+$380021.5 was estimated
to be 0.08 \citep{pav21j1727}, which seems to be consistent
with this interpretation.  There have been no indication
of WZ~Sge-type behavior neither in MASTER OT J055845.55$+$391533.4
nor MASTER OT J172758.09$+$380021.5, and these objects
may comprise a new class of rebrightening phenomenon in
SU UMa-type dwarf novae.
Further observations of MASTER OT J055845.55$+$391533.4 are
needed to establish the orbital period, and hopefully
the period of stage A to estimate $q$
\citep{kat13qfromstageA,kat22stageA}, and spectrocopy
is needed to estimate the chemical composition.
Since superoutbursts of MASTER OT J055845.55$+$391533.4 are
not rare, planned observations for a future superoutburst
are requested.

\section*{Acknowledgements}

This work was supported by JSPS KAKENHI Grant Number 21K03616.
The authors are grateful to the ATLAS, ZTF and ASAS-SN teams
for making their data available to the public.
We are grateful to VSOLJ and VSNET observers (particularly
Yutaka Maeda, Mitsutaka Hiraga and Masayuki Moriyama) who reported
snapshot CCD photometry of MASTER OT J055845.55$+$391533.4.
We are also grateful to Naoto Kojiguchi for helping downloading
the ZTF data and to Junpei Ito and Katsuki Muraoka for
converting the data reported to the VSNET Collaboration.

This work has made use of data from the Asteroid Terrestrial-impact
Last Alert System (ATLAS) project.
The ATLAS project is primarily funded to search for
near earth asteroids through NASA grants NN12AR55G, 80NSSC18K0284,
and 80NSSC18K1575; byproducts of the NEO search include images and
catalogs from the survey area. This work was partially funded by
Kepler/K2 grant J1944/80NSSC19K0112 and HST GO-15889, and STFC
grants ST/T000198/1 and ST/S006109/1. The ATLAS science products
have been made possible through the contributions of the University
of Hawaii Institute for Astronomy, the Queen's University Belfast, 
the Space Telescope Science Institute, the South African Astronomical
Observatory, and The Millennium Institute of Astrophysics (MAS), Chile.

Based on observations obtained with the Samuel Oschin 48-inch
Telescope at the Palomar Observatory as part of
the Zwicky Transient Facility project. ZTF is supported by
the National Science Foundation under Grant No. AST-1440341
and a collaboration including Caltech, IPAC, 
the Weizmann Institute for Science, the Oskar Klein Center
at Stockholm University, the University of Maryland,
the University of Washington, Deutsches Elektronen-Synchrotron
and Humboldt University, Los Alamos National Laboratories, 
the TANGO Consortium of Taiwan, the University of 
Wisconsin at Milwaukee, and Lawrence Berkeley National Laboratories.
Operations are conducted by COO, IPAC, and UW.

The ztfquery code was funded by the European Research Council
(ERC) under the European Union's Horizon 2020 research and 
innovation programme (grant agreement n$^{\circ}$759194
-- USNAC, PI: Rigault).

\section*{List of objects in this paper}
\xxinput{objlist.inc}

\section*{References}

We provide two forms of the references section (for ADS
and as published) so that the references can be easily
incorporated into ADS.

\newcommand{\noop}[1]{}\newcommand{\hyphalt}{-}

\renewcommand\refname{\textbf{References (for ADS)}}

\xxinput{j0558aph.bbl}

\renewcommand\refname{\textbf{References (as published)}}

\xxinput{j0558.bbl.vsolj}

\end{document}